%% file: main.tex
\documentclass[sigconf]{acmart}

\setcopyright{none}
\usepackage{graphicx} 
\usepackage{arydshln} 
\usepackage{subcaption}
\usepackage{tablefootnote}

\title{What Makes Programmers Laugh? Exploring the Submissions of the Subreddit r/ProgrammerHumor.}

\author{Miikka Kuutila}
\email{miikka.kuutila@dal.ca}
\affiliation{
\institution{Dalhousie University}
\department{Faculty of Computer Science}
\city{Halifax}
\country{Nova Scotia, Canada}}

\author{Leevi Rantala}
\email{leevi.rantala@oulu.fi}
\affiliation{
\institution{ University of Oulu}
\department{M3S}
\city{Oulu}
\country{Finland}}

\author{Junhao Li}
\email{junhao.li@oulu.fi}
\affiliation{
\institution{University of Oulu}
\department{UBICOMP}
\city{Oulu}
\country{Finland}}

\author{Simo Hosio}
\email{simo.hosio@oulu.fi}
\affiliation{
\institution{University of Oulu}
\department{UBICOMP}
\city{Oulu}
\country{Finland}}

\author{Mika Mäntylä}
\email{mika.mantyla@{helsinki.fi/oulu.fi}}
\affiliation{%
\institution{University of Helsinki /  Oulu}
\department{Dept of Computer Science / M3S}
\city{Helsinki / Oulu}
\country{Finland}}

\copyrightyear{2024}
\acmYear{2024}

\begin{document}




\begin{abstract}
\textbf{Background}: Humor is a fundamental part of human communication, with prior work linking positive humor in the workplace to positive outcomes, such as improved performance and job satisfaction. 
\textbf{Aims}: This study aims to investigate programming-related humor in a large social media community. 
\textbf{Methodology}: We collected 139,718 submissions from Reddit subreddit r/ProgrammerHumor. Both textual and image-based (memes) submissions were considered. The image data was processed with OCR to extract text from images for NLP analysis. Multiple regression models were built to investigate what makes submissions humorous.
Additionally, a random sample of 800 submissions was labeled by human annotators regarding their relation to theories of humor, suitability for the workplace, the need for programming knowledge to understand the submission, and whether images in image-based submissions added context to the submission.
\textbf{Results:} Our results indicate that predicting the humor of software developers is difficult. Our best regression model was able to explain only 10\% of the variance. However, statistically significant differences were observed between topics, submission times, and associated humor theories. Our analysis reveals that the highest submission scores are achieved by image-based submissions that are created during the winter months in the northern hemisphere, between 2-3pm UTC on weekends, which are distinctly related to superiority and incongruity theories of humor, and are about the topic of "Learning". 
\textbf{Conclusions:} Predicting humor with natural language processing methods is challenging. We discuss the benefits and inherent difficulties in assessing perceived humor of submissions, as well as possible avenues for future work. Additionally, our replication package should help future studies and can act as a joke repository for the software industry and education.

\end{abstract}
\maketitle

\input{introduction}
\input{background}
\input{methodology}
\input{results}

\input{discussion}
\input{conclusion.tex}

\bibliographystyle{ACM-Reference-Format}
\bibliography{sample-base}

\end{document}

%% file: introduction.tex
\section{Introduction}\label{sec:introduction}

Previously, natural language text produced by programmers has interested the scientific community in various ways, including the detection of sentiment in textual communication~\cite{kuutila2020chat}, self-admitted technical debt~\cite{huang2018identifying}, and design discussions~\cite{mahadi2022conclusion}. Similarly, linguistic analysis of software user behavior in discussion forums has also gained interest recently~\cite{hellman2022characterizing}. Quantitative studies have examined factors affecting software developers’ mood and productivity, such as listening to music while working~\cite{moster2022zone}, working times~\cite{claes2018programmers} and work well-being~\cite{kuutila2021individual}.

However, prior work on humor used by programmers has primarily focused on qualitative~\cite{friedman2003framework} or negative aspects~\cite{borsotti2022humor} of humor, without utilizing modern text-mining or quantitative methods, or examining large online communities. This is surprising, as it has been shown that happy developers solve problems better~\cite{graziotin2014happy}, and thus understanding what programmers find funny would be useful in fostering a more positive work environment or monitoring mood at workplace. Similarly, analyzing the peak engagement times for programming related humor can tell us when programmers engage with programming related humor, and thus, tells us more about the context of when programmers seek humorous content.

One place to find humor produced by programmers is the subreddit ProgrammerHumor\footnote{https://www.reddit.com/r/ProgrammerHumor/}, which describes itself as a place ``For anything funny related to programming and software development''. Reddit is one of the most popular social media platforms, with over 50 million daily users. As of this writing, r/ProgrammerHumor has over three million members, making it one of the largest programming-related online communities. As submissions are scored through users ``upvoting'' and ``downvoting'' by users, and both submissions and comments are accessible programmatically through APIs, this makes it a prime target for studying humor used by programmers. Thus, to investigate what programmers find funny, we formulate the following research questions:

\def\firstrq{How are individual words related to submission score?}
\def\secondrq{Are different topics identified through LDA clustering related to submission score?}
\def\thirdrq{What themes can be identified in the topics that are related to higher and lower submission scores?}
\def\fourthrq{How is the time a submission is made related to the submission score?}
\def\fifthrq{How are different factors (theories of humor, not suitable for work content and images which add context to the humor) represented in the submissions, and how do they relate to submission score?}
\renewcommand{\theenumi} {RQ\arabic{enumi}}
\begin{enumerate}
    \item \firstrq
    \item \secondrq
    \item \thirdrq
    \item \fourthrq
    \item \fifthrq
\end{enumerate}
\renewcommand{\theenumi}{\arabic{enumi}}


%% file: background.tex
\section{Background}\label{sec:background}

\subsection{Theories of humor}\label{sec:topic}
According to Shaw~\cite{shaw2010philosophy}, there are three main theories of humor: superiority, incongruity, and relief theories. These theories try to ``explain what it means for something to be humorous or what occurs when one perceives something to be humorous''.  

Bardon~\cite{bardon2005philosophy}, superiority theory is related to ``humor we find in comedy and in life is based on ridicule, wherein we regard the object of amusement as inferior and/or ourselves as superior.''. This theory is rooted in the works of Plato, Aristotle, and Hobbes, and has been widely discussed (e.g.,~\cite{lintott2016superiority, bardon2005philosophy}). This theory has informed studies in management sciences, such as one by Huang~\cite{huang2022leader}, where self-deprecating humor was found to positively affect leader identification.

Relief theory, another humor-related theory, suggests that laughter results from nervousness, excitement, or emotional tension, a concept attributed to Herbert Spencer~\cite{bardon2005philosophy}. From this point of view, humor in the workplace is used to the alleviate stress~\cite{wilkins2009humor}. Empirical evidence for the alleviation of stress has been established in the medical field~\cite{kimata2004laughter}.

Lastly, according to Bardon~\cite{bardon2005philosophy} the incongruity theory finds humor in the ``intellectual recognition of an absurd incongruity between conflicting ideas or experiences''. Thus, things that violate accepted patterns in a non-threatening way are perceived to be humorous. In other words, humor has an element of surprise or novelty in it. 

However, all of these theories have been criticized for that they ``rarely define their basic terms formally, and are insufficiently developed to make precise falsifiable predictions developed to make precise falsifiable predictions.''~\cite{ritchie2004linguistic}. A basic example of this given by Ritchie~\cite{ritchie2004linguistic}, who questions why some humor based on feelings of superiority is funny, while other instances with similar content related to feelings of superiority are not. A practical example of this is humor ridiculing a specific politician, which can be perceived as either funny or mean, even with the same target of ridicule.

\subsection{Humor in workplace}\label{sec:hum}

According to a meta-analysis by Mesmer et al.~\cite{mesmer2012meta}, humor in the workplace is associated with several positive effects, including enhanced work performance, satisfaction, work-group cohesion, health, and coping effectiveness, as well as decreased burnout, stress, and work withdrawal. Supervisor use of humor is associated with enhanced subordinate work performance, satisfaction, perception of supervisor performance, satisfaction with the supervisor, and work-group cohesion, as well as reduced work withdrawal.

However, jokes in the workplace can also be found offensive. What is found offensive can depend on things such as ethnicity, sex, or social power ~\cite{knegtmans2018impact, sacco2021not}. A review of disparaging humor by Ford et al.~\cite{ford2015disparagement} supports the hypothesis that disparagement humor targeting an out-group can foster prejudice toward that group. Ford et al.~\cite{ford2015disparagement} further suggest that disparagement humor primarily acts as a release of existing prejudice. At worst, offensive jokes can lead to litigation\cite{duncan1990humor}.

\subsection{Humor in Software Engineering Literature}\label{sec:humorse}
The earliest works that we know of that consider humor in computer science or software engineering context are by Linda Friedman and Hershey Friedman~\cite{friedman2002computer, friedman2003framework}. They study what they call computer oriented humor (Cohum), and note its anti-establishment and hermetic, knowledge-based nature. This humor is characterized as ``I-get-it'' humor, where the audience needs computer specific knowledge to understand it.

Recently, Borsotti and Bj{\o}rn~\cite{borsotti2022humor} investigated humor in computer science organizations, and notice that negative stereotypes create barriers for inclusion and creating a welcoming environment. As a result of analyzing humor used in the observed organizations, the authors propose three principles for organizations: 1) Examine organizational traditions and spaces to critically evaluate challenges for inclusion; 2) Normalize critical refection in the core practices of the organization; 3) Diversify and improve data collection.

Somewhat related work in the software engineering field has been done on toxicity detection. Sultana et al.~\cite{sultana2021rubric} has proposed a rubric for classifying sexist and misogynistic messages. Similarly, Sarker et al.~\cite{sarker2023toxispanse} developed a toxicity detector for code review comments.

\subsection{Natural Language Processing}\label{sec:humornlp}
Sentiment analysis considers sentiment polarity and distinct emotions. A recent systematic review and a mapping study of sentiment analysis in software engineering~\cite{obaidi2021development, obaidi2022sentiment} reveal the popularity of these NLP techniques. 

Recent work in humor detection outside of software engineering has achieved around 90\% accuracy in the ``Pun of the Day'' and one-liner datasets~\cite{FAN2020105}. The interpretability of the internal and external neural network (IEANN) by~\cite{FAN2020105} is improved with attention maps. 

Lin et al.~\cite{LI2020102290} added ``optimistic humorous'' and ``pessimistic humorous'' categories alongside traditional positive and negative sentiment polarity detection in their HEMOS system. Their system achieves F1-scores of 81.48\% for the optimistic humorous category and 68.42\% for pessimistic humorous categories in the context of Chinese social media posts. For longer texts, Morales and Zhai~\cite{morales2017identifying} propose a generative language model using supervised learning, achieving an accuracy of 86\%. 

\subsection{Topic modeling of Social Media}\label{sec:topic}
Early work on topic modeling in social media focused on establishing the methodology, comparing modeling results, demonstrating its usefulness in classification problems~\cite{hong2010empirical}, and using it as part of user recommendation systems~\cite{pennacchiotti2011investigating}.

Social media provides large, often unstructured datasets for analysis. As a result, topic modeling has been proposed for commercial purposes, such as product planning~\cite{JEONG2019280} and stock market prediction~\cite{nguyen2015topic}. The history of topic modeling dates back to 1990's~\cite{churchill2022evolution}. Recent work~\cite{thompson2020topic} has proposed using BERT~\cite{devlin2018bert} and LLM-based approaches for clustering~\cite{radford2019language} .

%% file: methodology.tex
\section{Methodology}\label{sec:methodology}

Submissions in the social media site Reddit are, by default, listed under the ``hot'' category, which shows recent submissions that have received many ``upvotes''. Users can also ``downvote'' submissions, which causes them to appear lower in the list of hot submissions. A submission's score is the number of upvotes minus the number of downvotes, although the score cannot go below 0. Additionally, the site uses score fuzzing\footnote{https://www.reddit.com/r/help/wiki/faq/}. Rediquette\footnote{https://www.reddit.com/wiki/reddiquette/}, which consists of written informal values by users of Reddit, states: ``If you think something contributes to conversation, upvote it. If you think it does not contribute to the subreddit it is posted in or is off-topic in a particular community, downvote it.''. Thus, total votes scores are related to how humorous submissions are perceived by the users of the topical subreddit ProgrammerHumor.

\subsection{Mining the Dataset}\label{sec:mining}

The Reddit submissions were acquired by mining the whole ProgrammerHumor subreddit using the third-party API Pushshift~\cite{baumgartner2020pushshift}. However, the Pushshift API saves the submissions very quickly after they are made and does not update their meta-data frequently. Thus, we used the python Reddit API wrapper (PRAW)\footnote{https://praw.readthedocs.io/} to update the metadata of the acquired submissions. The updated meta-data included a score that is the total score of the submission, the ratio of upvotes to downvotes and the number of comments the submission received.
 
The earliest submission in the acquired dataset is made in 2nd of January 2014 and the latest in 3rd of November 2022. We extracted time related information from the posts using the ''created\_utc'' attribute, which provides the submission creation time in UNIX time. To acquire the image based dataset, we took the submissions that were classified as non text-based submissions by the Reddit API's is\_self function, and used Googles Cloud Vision API\footnote{https://cloud.google.com/vision} for optical character recognition(OCR). Of the 203,799 submissions we sent to the API, we received recognized text to 129,668 image-based submissions. The OCR can fail for various reasons, but in our case this is most likely because the links to hosted images in the submissions were no longer accessible.

\subsection{Clustering}\label{sec:cluster}

Clustering was performed using Python version 3.10.12. For preprocessing the dataset, we used the Python library NLTK~\cite{bird2009natural}. The preprocessing steps included removing punctuation and numbers, converting text to lowercase, removing stopwords, and applying the library’s stemmer to transform words to their base form..

We used the python library sklearn and the function LatentDirichletAllocation to cluster the acquired submissions into topics. Latent Dirichlet Allocation (LDA) is a soft clustering algorithm ideal for text~\cite{blei2003latent}, allowing word stems to belong to multiple topics, and some documents may not belong to any topics. To find the optimal number of topics, we calculated perplexity scores for LDA models with the same package sklearn, and checked perplexity scores for the range between 15 and 75 topics. We selected the model with the lowest perplexity score, which was 23 topics. We share a graph showing perplexity score across the number of topics and the code for generating the topics in our replication package~\cite{zenodo_anon}. The ten most probable word stems for each identified topic are shown in the Table~\ref{tab:topicdetails}. The final dataset, which includes the most probable topic variable for each submission, contains 139,718 submissions, and it consists of both image and text based submissions.

\subsection{Examining the topics}
To assess both the worst and best performing topics in relation to submission scores, the first author examined sample submission with a focus on the most probable word stems for these topics. The goal was to understand how the submissions relate to the word stems. Additionally, the author noted any overarching themes observed during this process. For each topic under review, all manually labeled submissions and the 50 highest-scoring submissions were read (see Section~\ref{sec:label}). For the topic with the lowest coefficient, a random sample of 50 submissions with scores ranging between 10 and 200 was selected. Brief examples of jokes were also selected from each topic to illustrate the underlying themes.

\subsection{Bag of Words Model}
To investigate the first research question, we build a linear regression model with a lasso penalty using the function cv.glmnet from the R-package glmnet~\cite{friedman2021package} version 4.1.8. The model uses a document-term matrix made from the submission title and submission text to predict the submission score. We used 10-fold cross-validation when building the model. Additional preprocessing steps included converting text to lowercase and pruning, which involved removing words that appeared in the documents fewer than sixty times. Preprocessing was largely done with the package tm~\cite{feinerer2008text}, version 0.7.13. To explain the coefficients reported in the results section, the first author read submissions containing the word stem.

\subsection{Submission Labeling}\label{sec:label}

To gain a deeper understanding of the submissions, two of the authors labeled several aspects of each submission. Specifically, the labelers had a document where they marked whether a submission is related a) superiority theory, b) incongruity theory, c) to relief theory, or u) unknown or any combination of these three theories. If the humor in a submission was perceived to be related to multiple theories, the labeler noted which theories in a separate field. The authors also coded whether the submission was suitable for a work environment, whether programming knowledge was needed to understand the humor, and whether the image in image-based submissions contributed to the humor. If a labeler did not understand what was humorous about a submission, it was coded as unknown for the theory part. Generally, these unknown submissions were related to spam and were relatively few in number.

For this labeling task, we randomly selected 400 image-based submissions and 400 purely textual submissions. The document provided to the labelers included the submission title, any accompanying text, a link to view the full submission for context, and text extracted from images via OCR. After the first 100 textual submissions were labeled, the authors met and discussed the disagreements they had, to come to a better shared understanding of the labels. However the labels that had been done beforehand were not changed at this point. After this discussion, the remaining randomly sampled submissions were labeled, and inter-rater agreement was calculated. The inter-rater agreement between annotators is shown in Table~\ref{tab:labeling}. The Cohen's kappas are between 0.44 and 0.71, indicating moderate to substantial agreement according to the thresholds by Landis \& Koch~\cite{landis1977measurement}.

After completing labeling the random sample, the authors met and had several lengthy discussions on cases where there was disagreement. Here, we describe some of the general decisions made when labeling particularly difficult cases. We classified all submissions that we understood as self-deprecating humor as related to both superiority and relief theories, recognizing that the related theory could change depending on the reader's perspective. Similarly, many submissions were ambiguous and open to multiple interpretations. In such cases, we labeled the submission with multiple theories if no single interpretation seemed predominant. We also observed that similar humor could be classified differently with minor changes in wording. For example, submission mocking badly made ticket (superiority theory) versus recounting the experience of creating a flawed ticket (relief theory). Furthermore, we noted the significant impact of submission titles on interpretation, alongside the accompanying text and images, as they often dictated the intended perspective of the joke.

We labeled all submissions that included software development terminology as not understandable by non-developers. While non-developers might grasp the general gist of some of these submissions, such as a developer venting about their everyday work (relief theory), the details would likely be missed. Thus, for example, we labeled all submissions including references to stackoverflow, programming languages as not understood by non-developers. We also labeled all submissions related to math concepts related to programming (e.g. recursion, binary numbers and logic) as something non-developers would not understand. Although these math concepts are not strictly specific to software development, they are not commonly needed in most jobs.

When labeling whether a submission was safe for work, we considered whether it would be appropriate to tell the joke in a corporate office setting, rather than simply focusing on the words used. Thus submission with sexual content such as, for example, pickup-lines and double entendres, were marked as not safe for work. Other factors that led to a submission being labeled as not safe for work included references to drugs, religion, nationality, or ethnicity. We also marked submissions having violent content as not safe for work, with only exception being cartoons.

\begin{table*}
\caption{Labeling. For theory, a corresponds to superiority theory, b corresponds to incongruity theory, c to relief theory and u to unclear or multiple. For other categories y corresponds to yes and n to no.}
    \label{tab:labeling}
    \centering
\begin{tabular}{l|lll}
Label & Annotator1 & Annotator2 & Cohen's Kappa \\
   \hline
Image - Theory & a=122, b=77, c=92, u= 109 & a=147, b=71, c=71, u=111 & 0.54 \\\
Text - Theory & a=123, b=85, c=61, u= 131 & a=138, b=83, c=23, u=156 & 0.44 \\\
Image - NSFW & n = 341, y=59 & n = 347, y=53 & 0.71  \\\
Text - NSFW & n =351, y=49 & n=348, y= 52 & 0.69  \\\
Image - Comprehensible & n = 162, y=240& n = 196, y=204 & 0.58 \\\
Text - Comprehensible & n = 290, y=110 & n=234, y=166 & 0.58 \\
Image - Adds Context & y=323, n=77& y=278 n=122 & 0.60\\
        \hline
    \end{tabular}
\end{table*}

\begin{table}[!htb]
\caption{Consensus Labels.}
\resizebox{\linewidth}{!}{%
\begin{tabular}{l|ll}
Data Label & Counts & \% \\
\hline
Image Theory & Superiority - 121 & 30\% \\
        & Incongruity -  75  & 19\% \\
        & Relief - 79  & 20\% \\
       & Unknown, multiple- 125 & 31\%\\
        & Multiple - all 36 & 9\%\\
        & Multiple - a and b  18 & 5\% \\
        & Multiple - a and c  35 & 9\%\\
        & Multiple - b and c  15 & 4\%\\
        & Not understood 17 & 4\%\\
        & Not available - 4 & 1\%\\
\hdashline
Text Theory & Superiority - 121 &  30\% \\
       & Incongruity -  94  &  24\%\\
       & Relief - 35  &  9\%\\
       & Unknown or multiple total- 150 & 38\%\\
       & Multiple - all 17 & 4\%\\
        & Multiple - a and c 29  & 7\%\\
        & Multiple - a and b 34 & 9\%\\
        & Multiple - b and c 10  & 3\%\\
        & Not understood or spam 59 & 15\% \\
       & Deleted - 1 & 0\%\\
\hline
Image  NSFW & Yes - 67 & 17\%\\
   & No 333 & 83\%\\
\hdashline
Text  NSFW & Yes - 58 & 15\%\\
   & No 342 & 86\%\\
\hline
Image  Comprehens. & Yes - 152 & 38\%\\
   & No 248 & 62\%\\
\hdashline
Text  Comprehens.& Yes - 121 & 30\%\\
   & No 279 & 70\%\\
\hline
Image Adds & Yes - 320 & 80\%\\
Context   & No 80 & 20\%\\
\hline
    \end{tabular}
}
\end{table}

\subsection{Statistical analysis}\label{sec:stat}

All statistical analysis were performed using the R programming language version 4.3.3. For fitting regression models we used the function glm from the package stats, which is part of base R. In this paper, we fitted three regression models: first one predicting submission score with individual word stems (Table~\ref{tab:glmnet}), another predicting submission score with topics identified with clustering and the time of the day the submission was made (Tables~\ref{tab:topicsregression} and ~\ref{tab:timeregression}), and a third model predicting submission score with human annotated labels (Table~\ref{tab:labelregression}). The second model is presented in two tables due to a large number of coefficients. Inter-rated reliability statistic, Cohen's kappa, was calculated using the package irr~\cite{gamer2012package} version 0.84.1. Our dataset, human labeling, and analysis scripts are available in our Zenodo replication package~\cite{zenodo_anon}.

%% file: results.tex
\section{Results}\label{sec:results}

\subsection{\textbf{RQ1: \firstrq}}\label{sec:RQ1}
Table~\ref{tab:glmnet} shows the ten highest positive and negative coefficients for the lasso regression model. Overall the $R^2$ for the model is 0.10 $R^2$, with a mean squared error of 0.95. Most of the positive coefficients are related to creators of humor, partly because these creators sign the images they produce, which then circulate in the subreddit submissions and are detected by the OCR software.

Many positive coefficients for the model are related to cartoons. The word ``monkeyusercom'' is a website hosting programming-related cartoons by a single creator. Similarly, the word ``toon'' comes from the logo of ``webtoons.com'', a host of different web comics, whose logo is attached to comic strips to direct people to more of their work. The list of positive coefficients also includes another cartoon website ``commitstripcom'' and the word stem ``endang'' which refers to the web cartoon ``Safely Endangered''. The word stem ``yuvakrishnamem'' refers to a user who self-describedly ``creates memes'', and who embeds a link to their humoristic images with a link to their Facebook group with around 60,000 members.

The positive coefficients refer to two different twitter/X accounts. Two coefficients ``iamdevlop'' and ``devlop'' are related to twitter/X user ``@iamdevloper'', who is a humoristic Twitter/X user with half a million followers. Submissions with these word stems typically include screenshots of their humorous tweets. The most positive word stem ``pathol'', and the word stem ``ppathol'' are related to screenshots of tweets by Twitter/X user ``@PPathole'', who has almost 200k followers at the time of writing. Similarly, the word ``retweet'' appears in the screenshots of humoristic tweets shared in the subreddit.

The most negative coefficient is ``mrw'', an abbreviation for ``My reaction when''. These submissions typically consist of a title that describes what the submitter did and their reaction, such as ``MRW I realized all my errors were caused by a missing ;'', followed by a picture of a crying person. The abbreviations ``mfw'' ('My face when') and ``tfw'' ('That feeling when') appear in submissions similar in content to those with ``mrw''.

The word stem ``imgurcom'' comes from submissions with the text ``The image you are requesting does not exist or is no longer available. imgur.com'', which the OCR returns for images that have been deleted from this image-host. The word stem ``upcom'' is usually related to submissions that have the word ``upcoming''. These submissions include questions that do not appear to have humorous content and therefore do not fit the subreddit, such as ``How do I prepare for an upcoming job interview'', news articles and spam.

The word stem ``airbnb'' is mostly associated with screenshots of a push notification, apparently sent to airbnb-users, which simply reads ``Test test dev''. The majority of these submissions feature the same image, which is resubmitted repeatedly. The word stem ``isevenint'' is contained in submissions that have either pseudo code, or programmatic code that has unnecessarily complicated method, which checks whether a given number is even. This could be for example a never-ending switch case. The word ``stackstatus'' is linked to submissions featuring a screenshot of ``stackoverflow.com'' being offline, with titles like ``This is my nightmare''. The word stem ``rli'' comes from typing the word really as ``rly'', and again the vast majority of submissions containing this word stem are screenshots of a single tweet which reads: ``Every time I have a programming question and I rly need help, I post it on Reddit and then log into another account and reply to it with an obscenely incorrect answer. Ppl don't care about helping others but they LOVE correcting others. Works 100\% of the time''. Lastly, the word stem ``artifact'' comes from a meme image that references the FOSS community, where people edit the image slightly and repost it to the subreddit. The editing typically involves fixing typos while adding new ones, and the image is accompanied by text such as ``When your pull request that fixed a small typo is accepted and you see your name among the Github contributors''. The common theme for these coefficients is that many submissions are identical or nearly identical in content, with few submissions with an earlier time stamp having high scores while many later submissions have a score of 0.


\begin{table}[!htb]
\caption{Lasso regression model explaining submission score with uni-grams, only the top 10 positive and negative coefficients shown. N is the number of submissions with at least a single occurance of the word stem.}
\label{tab:glmnet}
\centering
\begin{tabular}{lll|lll}
\hline
Positive Coef & Est & N & Negative Coef & Est & N\\
\hline
pathol& 0.02 & 91& mrw& -0.03 & 169\\
monkeyusercom& 0.02 &279 & mfw & -0.02& 65\\
toon& 0.02 & 100& airbnb& -0.02& 99\\
devlop& 0.02& 354& imgurcom& -0.02& 582\\
ppathol& 0.02 & 90& upcom& -0.02& 61\\
iamdevlop& 0.02& 291& tfw& -0.02& 87\\
yuvakrishnamem& 0.02 & 312& isevenint& -0.02& 98\\
commitstripcom& 0.02& 186& stackstatus& -0.02& 161\\
retweet& 0.02& 1619& rli& -0.01& 72\\
endang& 0.02 & 75& artifact& -0.01& 58\\
\hline
 $R^2$:0.10 & && MSE: 0.95 & \\
\hline
\end{tabular}

\end{table}

\subsection{\textbf{RQ2 - \secondrq}}\label{sec:RQ2}

\begin{table*}
\caption{Number of submissions for each topic, and the most probable word stems for each topic.}
    \label{tab:topicdetails}
 \centering
\begin{tabular}{l|llll}
Topic & 10 most probable word stems & N & Mean Score & Mean Comments\\
\hline
Topic 1:& div, css, id, fuck, class, password, html, text, style, script & 4766 & 680.68 & 24.00\\
Topic 2:& data, key, ctrl, null, vim, use, sep, mar, exit, databas & 4050 & 853.13 & 30.80 \\
Topic 3:& self, array, start, const, count, step, tabl, bit, row, word & 2785 &731.25 & 23.81 \\
Topic 4:& error, line, commit, fix, code, space, minut, end, nan, compil & 5366 & 837.74 & 21.51 \\
Topic 5:& learn, comput, googl, machin, recurs, android, studio, hack, visual, chicken & 3164 &1426.50 & 36.37\\
Topic 6:& code, bug, pm, work, today, fix, user, github, like, know & 11317 & 1154.74 & 24.58\\
Topic 7:& python, programm, html, use, java, real, program, root, json, youtub & 5992 & 1078.80 & 40.11\\
Topic 8:& ago, comment, repli, share, hour, save, post, like, report, point & 6065 & 1824.50 & 42.50\\
Topic 9:& need, time, want, use, work, know, like, help, ll, order & 14329 & 980.39 & 30.70\\
Topic 10:& com, imgflip, stack, overflow, question, project, answer, stackoverflow, strong, ask &7967 & 1015.14& 28.56\\
Topic 11:& window, run, type, doo, app, instal, linux, microsoft, help, command & 5441 &746.83 &27.17\\
Topic 12:& defin, case, main, includ, break, day, int, std, variabl, build & 2427 & 878.37 &35.30\\
Topic 13:& javascript, number, log, object, js, man, console, ai, engin, phone & 3115 & 830.57 & 30.37\\
Topic 14:& return, true, function, els, var, fals, int, number, loop, false & 4423 & 716.25 & 32.41 \\
Topic 15:& develop, web, experi, year, softwar, design, job, manag, mb, video & 8295 & 1006.43 & 34.23\\
Topic 16:& make, thing, ye, problem, let, say, someth, know, com, tri & 10327 &1053.96 & 30.67\\
Topic 17:& like, good, guy, joke, feel, look, know, peopl, programm, im & 8871 &1036.83 &33.01\\
Topic 18:& com, www, https, http, page, org, ms, reddit, search, server & 6102 & 536.33 & 37.05\\
Topic 19:& java, string, public, int, class, void, static, new, main, privat & 3724 & 614.82 & 30.67\\
Topic 20:& test, dev, php, unit, pass, vs, senior, backend, png, undefin &3414 & 918.53& 23.78\\
Topic 21:& file, git, list, new, date, game, email, add, master, request & 6205& 704.56& 23.45\\
Topic 22:& program, code, use, languag, write, think, compil, chang, read, cs & 9618 &1068.63 &39.00\\
Topic 23:& world, hello, print, import, num, pi, elif, node, path, random & 1955&956.16&36.35\\
   \hline
    \end{tabular}
\end{table*}

Coefficients for the regression model predicting submission score with topics are shown in Table~\ref{tab:topicsregression}. Overall, the different topics identified with clustering do not predict submission score well. The $R^2$ value for the overall model is 0.081. In addition to topics, the model includes time related coefficients (shown in Table ~\ref{tab:timeregression}, whether a submission is image- or text-based, and the lengths of the title and submission text. However, the topics do differ from each other in a statistically significant way, and overall, the predictor variables do add prediction power to the model in a statistically significant way over a null model, as evidenced by the p-value of the F-statistic. We also added information on whether the submission was image or text based according to the Reddit API, and the character lengths of title of the submission and the text that was part of the submission (written or the result of OCR).

We chose Topic 1 to be the reference category to which other topics are compared to out of convenience, because there was no theoretical basis to assume any differences between topics. The five highest coefficients are achieved by Topics 5, 22, 6, 8 and 16. On the other hand, the five worst coefficients for topics are by Topic 18, 20, 19, 1 and 3. The differences between them and the topics with highest coefficients are all statistically significant, as evidenced by the extremely high p-values (< 2e-16).

\begin{table}[!htb]
\caption{Linear regression predicting submission score - Topics, image or text based submission and the model effect size and significance measures. Same model continued in Table 6.}
\label{tab:topicsregression}
\resizebox{\linewidth}{!}{%
\begin{tabular}{l|llll}
Variable & coef & std.err & t  & p-value  \\
\hline
Reference: &  Topic - 1\\
Topic - 2   &   0.124  & 0.02  & 6.18 & 6.29e-10 *** \\
Topic - 3  &   0.024  & 0.02  &  1.07 & 0.283963  \\
Topic - 4  &   0.131  & 0.02   & 7.016 & 2.30e-12 *** \\
Topic - 5   &   0.260  & 0.02 & 12.07  & < 2e-16 *** \\
Topic - 6   &   0.217   & 0.02  & 13.31  & < 2e-16 *** \\
Topic - 7   &   0.175 & 0.02   & 9.59  & < 2e-16 *** \\
Topic - 8   &   0.208  & 0.02  & 11.43 &  < 2e-16 *** \\
Topic - 9   &   0.185  & 0.02   & 11.70 & < 2e-16 *** \\
Topic - 10  &   0.155  & 0.02  & 8.95 & < 2e-16 *** \\
Topic - 11  &   0.046  & 0.01   & 2.47  & 0.012904 *   \\
Topic - 12  &   0.113 & 0.02  & 4.85 & 1.25e-06 *** \\
Topic - 13  &   0.061  & 0.02   & 2.83 & 0.004612 ** \\
Topic - 14  &   0.052 & 0.02   & 2.66 & 0.007931 **  \\
Topic - 15  &   0.123 & 0.02   & 7.20 & 6.18e-13 ***\\
Topic - 16  &   0.182  & 0.02   & 11.02  & < 2e-16 ***\\
Topic - 17  &   0.178 & 0.02  & 10.53 & < 2e-16 ***\\
Topic - 18  &  -0.083 & 0.02  & -4.55  & 5.40e-06 ***\\
Topic - 19  &  -0.002 & 0.02 & -0.08 & 0.939104  \\
Topic - 20  &  -0.008  & 0.02  & -0.40  & 0.691577 \\
Topic - 21  &  0.035 & 0.02  &  1.98 & 0.056398 .   \\
Topic - 22  &   0.227  & 0.02 & 13.58 & < 2e-16 ***\\
Topic - 23  &   0.077  & 0.03  &  3.06 & 0.002225 **  \\
\hdashline
Reference & Image &-Based  & \\
Text-Based  & -0.768 &  0.01 & -76.508 & < 2e-16 *** \\
\hdashline
Title Length & < -0.01 & <0.01 & -28.035 & < 2e-16 *** \\
Text Length & < -0.01 & <0.01  & -9.595 & < 2e-16 *** \\
\hline
 $R^2$:0.081 &F-Stat &p-value &< 2.2e-16*** \\
\hline
\end{tabular}
}
\end{table}

\subsection{\textbf{RQ3 - \thirdrq}}\label{sec:RQ3}
Here we give a brief overview of the topics that had the 3 highest coefficients when predicting submission scores, that is topics 5, 22 and 6, and the lowest coefficient, that is topic 18, when predicting submission scores.

\subsubsection{Topic 5 - Learning.}
The top-scoring submissions in this topic generally related to four themes, all connected to learning in some way. First, there is a self deprecating irony in many submissions, such as comparing learning by googling between doctors and software engineers, or commentary on how learning more about science or software engineering affects trust in the results. An example of this theme is a submission titled ``Shit, they are onto us!'' featuring an image of two coffee mugs for sale: one reads, ``Please don't confuse your Google search with my medical degree,'' and the other says, ``Please don't confuse your Google search with my Google search''. Second, many submission made fun of machine learning. An example of this would be a comparisons checklist between machine learning algorithm and a parrot, where items such as  ``Learns random phrases'',  ``Doesn't understand shit about what it learns'' and ``Occasionally speaks nonsense'' are checked for both. Third, many submissions were related to learning to use visual and android studio and their perceived shortcomings, such as time it takes to start them or memory usage. Lastly, some submissions focused on the difficulty of learning math-related concepts for hobby projects, such as recursion or discrete math.

\subsubsection{Topic 22 - Writing Code}
The top-rated submissions from this topic were almost always related to writing code in some way. This included stories about close relatives giving vague requirements for writing a program or self-deprecating tales of writing a slower sorting algorithm in C compared to a professor's implementation in Python. Many submissions were related to particular programming languages, e.g. marveling at programs written in Assembly or stories about selecting programming languages based on very arbitrary qualities. One submission was repeated multiple times in the top submissions, that is a screenshot of a submission from the main programming subreddit focusing on discussing computer programming in a serious manner, which shows a link to a blog entry about programmers being ``probably wrong'' in wanting to delete old code, which is further encapsulated in the statement ``fundamental law of programming: It's harder to read code than to write it''. However, the most upvoted comment to that serious blog entry is ``Also the old code is a mess.''.

\subsubsection{Topic 6 - Fixing Bugs}
A unifying theme in this topic was stories and observations related to fixing bugs. These could be references to trying to find fixes to specific obscure problems, only to come up short in a particularly frustrating way. For example, a supposed fix might be behind a broken link, or someone might have asked a question related to the problem years ago with a single reply from the author saying, ``NM fixed it''. 

\subsubsection{Topic 18 - Links and servers}
This was the lowest-rated topic. Many of the top-rated submissions for this topic were related to problems with web servers, such as "404 errors" and commits that broke servers. Among the lower-voted submissions, the overarching theme was the inclusion of web addresses, often related to stack traces from web programming.

\subsection{\textbf{RQ4 - \fourthrq}}\label{sec:RQ4}

To analyze the peak engagement times for programming related humor, we also added time related variables to the model shown in Section~\ref{sec:RQ2}.The time related coefficients, which are shown in Table~\ref{tab:timeregression} refer to time in UTC. The $R^2$ for the model is 0.081, meaning that the time-related coefficients, along with topics, title and text lengths, and whether a submission is image- or text-based, explain 8.1\% of the variance. In general, the coefficients for year and hour of the day vary more than the coefficients for weekday and month.

As shown, posts made on the weekend, particularly Saturday and Sunday, have positive coefficients compared to Mondays. Other weekdays have worse coefficients than Monday, with Wednesday having the worst coefficient. Perhaps surprisingly, Friday in UTC is a worse day than Monday in UTC to create a submission.

The reference category for the year is 2014, which is the first year in our dataset. Based on our model, submission scores were lower in 2015 and 2016, but then steadily rose until 2021. The last year in the dataset, 2022, shows lower submission scores than 2021. For months of the year, the winter months of northern hemisphere in October, November, December and January are related to higher scores, where as the lowest are seen for the months of March, April and August.

In our model, the reference category for the submission hour is the first hour of each day in UTC. In general, the hours between 6am and 4pm UTC produce the highest coefficients for submission score. The highest submission score is achieved between 2pm and 3pm, after which the coefficients submission score starts to lower.

\begin{table}
\caption{Coefficients for predicting submission score - time related variables. Continued from Table 5.}
\label{tab:timeregression}
\centering
\resizebox{\linewidth}{!}{%
\begin{tabular}{l|llll}
Variable & coef & std.err & t  & p-value  \\
\hline
Reference: & Monday\\
Tuesday  &  -0.037 & 0.01& -4.02 & 5.75e-05 *** \\
Wednesday  &  -0.069 & 0.01&-7.46 & 8.53e-14 *** \\
Thursday &   -0.043 & 0.01& -4.67& 3.04e-06 *** \\
Friday & -0.044 & 0.01 & -4.77& 1.85e-06 *** \\
Saturday  &   0.040 & 0.01 &  4.10 & 4.11e-05 *** \\
Sunday  &   0.070 & 0.01 &  7.04 &  1.88e-12 *** \\
\hdashline
Reference & 2014 \\
2015  &  -0.142& 0.03 & -4.54 & 5.59e-06 ***  \\
2016  &  -0.178 & 0.03 & -5.71 & 1.15e-08 *** \\
2017  &  0.103 & 0.03 & 3.56 & 0.000366 *** \\
2018  &   0.076 & 0.03 &  2.69 & 0.007065 ** \\
2019  &  0.024 & 0.03 &  0.854 & 0.392973 \\
2020  &   0.040 & 0.03 &  1.42 & 0.154844   \\
2021  &   0.141 & 0.03 &  5.03 & 4.99e-07 *** \\
2022 &  0.047 & 0.03 &  1.67 & 0.094505 . \\
\hdashline
Reference & January \\
February  &    -0.048 & 0.01 & -3.80 & 0.000144 *** \\
March   &   -0.079 & 0.01 & -6.38 & 1.80e-10 *** \\
April   &   -0.072 & 0.01 & -5.83 & 5.62e-09 *** \\
May   &   -0.046 & 0.01 & -3.77 & 0.000162 ***  \\
June  &   -0.063 & 0.01 & -5.12 & 3.03e-07 *** \\
July   &   -0.074 & 0.01 & -6.08& 1.25e-09 *** \\
August  &   -0.077 & 0.01 & -6.24 & 4.47e-10 *** \\
September   &   -0.017 & 0.01 & -1.39 & 0.164969 \\
October   &   -0.005 & 0.01&  -0.38 & 0.706288   \\
November   &   0.024 & 0.01 &  1.87 & 0.062185 . \\
December   &   0.008 & 0.01 &  0.64 & 0.524162 \\
\hdashline
Reference & Hour 0-1 \\
Hour 1-2     &  -0.036 & 0.02 & -2.08 & 0.037851 * \\
Hour 2-3     &  -0.063 & 0.02 & -3.47 & 0.000548 *** \\
Hour 3-4     &  -0.044 & 0.02 & -2.33 & 0.019658 *  \\
Hour 4-5     &  -0.002 & 0.02& -0.13& 0.895205  \\
Hour 5-6    &   0.019 & 0.02 &  1.01 & 0.312794   \\
Hour 6-7     &  0.082 & 0.02 &  4.28 & 1.86e-05 *** \\
Hour 7-8     &  0.144 & 0.02 &  7.59 & 3.24e-14 *** \\
Hour 8-9      &  0.168 & 0.02 &  8.78 &  < 2e-16 *** \\
Hour 9-10     &   0.180 & 0.02 &  9.64 & < 2e-16 *** \\
Hour 10-11    &   0.179 & 0.02 & 9.82 & < 2e-16 *** \\
Hour 11-12    &   0.188 & 0.02 & 10.47 & < 2e-16 *** \\
Hour 12-13    &   0.191 & 0.02 & 10.82 & < 2e-16 *** \\
Hour 13-14    &   0.197 & 0.02 & 11.27 & < 2e-16 *** \\
Hour 14-15    &   0.220 & 0.02& 12.74 & < 2e-16 *** \\
Hour 15-16    &   0.170 & 0.02 & 10.22 & < 2e-16 *** \\
Hour 16-17    &   0.108 & 0.02& 6.62 & 3.60e-11 *** \\
Hour 17-18    &   0.103 & 0.02 &  6.64 & 1.29e-10 *** \\
Hour 18-19   &    0.106 & 0.02&  6.67 & 2.54e-11 *** \\
Hour 19-20   &    0.068 & 0.02 &  4.30 & 1.71e-05 *** \\
Hour 20-21   &    0.058 & 0.02 &  3.65 & 0.000262 ***\\
Hour 21-22   &    0.031 & 0.02 &  1.93 & 0.054245 .   \\
Hour 22-23   &    0.043 & 0.02 &  2.70 & 0.006924 **\\
Hour 23-24  &   0.002 & 0.02 &  0.14 & 0.888733  \\
\hline
\end{tabular}
}
\end{table}

\subsection{\textbf{RQ5 - \fifthrq}}\label{sec:RQ5}
The linear regression model using the labeling data is shown in Table~\ref{tab:labelregression}. All submissions related to theories have a positive coefficient compared to submissions that the labelers did not understand or identified as spam. Unsurprisingly, submissions that were not fully available had a negative coefficient compared to other submissions. The most positive coefficient is held by combination of superiority theory and incongruity theory, followed by incongruity theory and superiority theory individually. Lastly, the combination of superiority theory and relief theory is positively related to submission scores compared to submissions that were not understood by the labelers.

Additionally, submissions that had not safe for work content had lower scores than those that without, and this difference is almost statistically significant at p < 0.05 level. Submissions that required software engineering knowledge to be understood had a slightly negative coefficient compared to submissions that did not require this knowledge. Lastly, image-based submissions where the image did not add context had a higher score than submissions without images, or where the image gave context to the joke. 

\begin{table}[!htb]
\caption{Linear regression predicting submission score with various aspects of human labeling. For theory, a corresponds to superiority theory, b corresponds to incongruity theory, c to relief theory, and multiples to combinations of them. }
\label{tab:labelregression}
\centering
\resizebox{\linewidth}{!}{%
\begin{tabular}{l|llll}
Variable & coef & std.err & t  & p-value  \\
\hline
Reference: Spam or &   \\
Not Understood &   \\
Theory - a       &       0.34 &    0.11 &   3.03 &  0.00249 **  \\
Theory - a,b    &        0.53  &   0.15 &   3.46&  0.00056 *** \\
Theory - a,b,c   &       0.25 &    0.15  &  1.64 &   0.10   \\
Theory - a,c     &       0.18 &   0.15  &  1.24 &  0.22  \\
Theory - b       &       0.42  &   0.12  &  3.54 & 0.00043 *** \\
Theory - b,c     &       0.26 &   0.20  &  1.31 &  0.19    \\
Theory - c       &       0.23  &  0.13 &   1.79 & 0.074 .    \\
Theory - not available & -0.09  &   0.39&   -0.22 &  0.83    \\
\hdashline
Reference: NSFW - No &   \\
NSFW - Yes         &         -0.11 &    0.08 &  -1.36 &  0.175 \\ 
\hdashline
Comprehensible - No &   \\
Comprehensible - Yes     &    -0.07 &    0.07  &  -1.05 &  0.30  \\  
\hdashline
Adds Context- No &   \\
Adds Context - No image    &      -0.19  &   0.11 &  -1.77 &  0.077 .   \\ 
Adds Context- Yes            &      -0.14  &   0.11 &  -1.24 &  0.22  \\
\hline
$R^2$:0.033 & F-Stat & p-value  & = 0.0087 & \\
\hline
    \end{tabular}
}
\end{table}

%% file: discussion.tex
\section{Discussion}\label{sec:discussion}

In general practice, understanding humor in online social media has clear utility (see e.g.,~\cite{liu2021examining}). Understanding programmer humor with an NLP-based pipeline that produces real-time information about the topics that given communities find entertaining—and therefore shareable and potentially viral—would benefit advertisers and community managers. User engagement is critical in these contexts. Second, since positive humor has been linked to positive outcomes in the workplace (see e.g.~\cite{mesmer2012meta}), understanding and measuring humor from workplace communications could potentially provide a measure related to the mood in software development teams and projects. Third, as positive humor is related to positive outcomes, creating interventions with positive humor for developers and testing their effects on, for example, productivity and mood, would be valuable. Our work is a novel and modest step for further studies in this area. 

From the performance measures our models, it can be said that predicting what is seen as funny by developers is quite hard. Based on the $R^2$ scores, individual words explain about 10\% or the variance in the large dataset, where as factors such as whether submission is image or text based, title and submission text lengths, which LDA-topic it contains and when the submission was created explain 8.1\% of the variance. For the smaller dataset of human labeled data, the model explained only 3.3\% of the variance. These results are in line with some previous work~\cite{morales2017identifying, weissburg2022judging}. 

What can we say about what programmers find humorous then? The bag-of-words Lasso-regression model in Table~\ref{tab:glmnet} suggests that submissions from specific creators of humor are seen as more humorous than submissions on average. A lot of the content related to these word stems are single strip cartoons, and tweets by specific Twitter/X accounts that focus on programming related humor. Conversely, the submissions associated with negative coefficients are related to simple reaction pictures, frequently reposted submissions, and unavailable content.

The mean submission scores for different topics varied from a low of 536.33 to the high of 1824.50, as seen on Table~\ref{tab:topicdetails}. Thus, it is evident that while the topic a particular submission belongs to does not necessarily predict its submission score, some topics were seen in general more humorous than others. The brief overview of the top submissions in the highly rated topics in Section~\ref{sec:RQ3} tells us that generic actions of developers such as learning by googling, writing code and fixing bugs were associated with higher submission scores. This is unsurprising, as these are shared experiences among developers with which many can identify. On the contrary, submission with a long stack trace does not seem to amuse the average programmer. As noted in Section~\ref{sec:RQ1}, when examining submissions with certain word stems, many of the negative coefficients are initially found in highly scored submissions, which are later found in similar or identical submissions with low scores, indicating that the novelty of successful submissions diminishes over time. Additionally, image-based submissions, as categorized by the Reddit API, are more likely to have higher scores. Lower title and submission text lengths were also associated with higher scores. Thus, it seems that submissions that are novel, image-based, and less complex are more likely to score higher. However, submissions cannot be too simplistic, as evidenced by the negative coefficients of reaction submissions ``mrw'', ``mfw'' and ``tfw''in Section~\ref{sec:RQ1}.

Regarding user engagement, the highest submission scores were achieved by submissions created in the winter months of the northern hemisphere, between 2-3 p.m. UTC on weekends. As Reddit users in general are from western countries\footnote{https://backlinko.com/reddit-users}, it seems that programmers are more likely to seek humorous content related to programming on their free time in the evenings and weekends, and during cold months, instead of during working hours or summer months. The time related coefficients in the model presented in Table~\ref{tab:timeregression} intuitively make sense. For example, the number of subreddit users has grown over the years\footnote{https://subredditstats.com/r/ProgrammerHumor}, and thus, there are more people to ``upvote'' popular submissions. A confounding factor for the weekday variables is that submissions made after 4 or 5 p.m. on the West Coast of the United States (depending on daylight saving time) are part of the next day’s submission scores in our model. This likely shows in the coefficients for Friday and Monday. Hours in the relatively early morning UTC time gather the highest scores. This is likely because the scores from previous days submissions have been lowered for the algorithm that displays ``hot'' submissions of the day. Additionally, the new submissions that do make the ``hot'' submissions list are displayed to visiting users throughout the whole day.

For the data annotated by human labelers, submissions linked to theories of humor had higher scores than those that were not understood or were interpreted as spam. Particularly high coefficients were for incongruity theory and the combination of superiority and incongruity theory.

In our labeled data not safe for work content was not related to higher submission scores, though around 15\% of image-based labeled submissions did have not safe work content. This is somewhat encouraging result in the light of prior results~\cite{borsotti2022humor}, as it suggests that NSFW content is not rated higher than safe for work content. However, it should be noted that in our labeling process, submissions labeled ``nsfw'' were not necessarily ``toxic'', as they were not directed to anyone in particular. For example, a joke about recursion written into a pick-up line is not necessarily directed at anyone in particular at a social media context, where as telling the same joke directly at someone in a work setting would most likely be inappropriate.

The majority of the labeled submissions were not properly comprehensible to people without some programming knowledge, with 62\% of the image based submissions and 70\% text based submissions being labeled as requiring programming knowledge. Thus, the subreddit ProgrammerHumor could be understood as a software repository in a certain sense. However, whether a submission needed some programming knowledge did not predict submissions score. Thus, it seems that the ``I-get-it'' humor noted by Friedman and Friedman~\cite{friedman2002computer} is not particularly popular in the age of social media.

There are several factors that make predicting the score of humorous submissions difficult. First, theories of humor address several aspects that are subjective and thus hard to measure. For example, in relation to superiority theory, who or what is it appropriate to feel superior to? In relation to incongruity theory, what are accepted patterns and what are novel and surprising ways to break them?

Second, upvotes and downvotes are used by Reddit algorithms to display submissions. In practice this means a kind of ``winner takes all -effect'', where submissions which get lots of upvotes early are shown to more users, which then drives engagement and increases the potential submission score. In other works this has also been called the herding effect~\cite{weninger2015random}. This is clearly demonstrated by the fact that if the number of comments or upvote ratio is added as a predictor variable to the model shown in Tables~\ref{tab:topicsregression} and ~\ref{tab:timeregression}, the model explains 6\% and 27\% of the variance respectively. Third, the result is a highly skewed distribution, as evidenced in our dataset by a mean of 973 and a median of 73. We also want to note that our work is fundamentally different than the work in Section~\ref{sec:humornlp}, because we are modeling how much users of the programmerhumor subreddit ``upvote'' a given submission and the work in Section~\ref{sec:humornlp} is about classifiers which classify whether a given text is humor or not. One could argue that predicting whether something is humorous is easier than predicting whether it is good humor. This would make the prediction task in this paper more difficult compared to prior work. 

\section{Threats to validity and future work}
\subsection{Limitations}

As for limitations, this study focuses on a single, albeit large, topical community of programmers on the Internet. Thus the results are quite generic, and the humor used in specific smaller work communities might differ widely. More work could be done by focusing on specific work communities, which could have their own ways of using humor related to their work context and culture.

One could also add an additional similar subreddits to the analysis, for example, the subreddit ProgrammingHumor in addition to ProgrammerHumor from the same social media site. However, we believe the benefits of this would be quite limited, because there is a widely used phenomenon of crossposting in reddit\footnote{https://reddit.zendesk.com/hc/en-us/articles/4835584113684-What-is-Crossposting-}, where submissions from one subreddit are shared from one community to other similar ones. Similarly, additional communities with similar content can be found in social media sites such as Facebook, however, they have different algorithms for visibility and scoring mechanisms, thus likely making prediction more difficult.

There is a significant survival threat in our data. Because the data is mined years after the original submissions have been made, and we only have access to submissions that were not deleted when the OCR was run. This means the older a particular submission is, the higher the chance is that its no longer available. 

When familiarizing with the dataset, we noticed that there are plenty of duplicate submissions or submissions that differ in a very minimal way from one another. These can be either posted very close to each other or sometimes many years apart, and the score these submissions have received might vary widely as well. However, the APIs we used do not store any information on when the upvotes are received, and we do not know of a way to get this information. Building a dataset that has information on how quickly upvotes are received, lets say the first hour, might control better for the aforementioned herding effect and lead to less skewed score distribution to analyze.

\subsection{Future Work}

Future work could include investigating who the users of the subreddit ProgrammerHumor are and their reasons for using the subreddit through surveys. However, third-party website\footnote{https://subredditstats.com/r/programmerhumor} indicates that the subreddit has grown at an exponential rate since 2018, and most users do not comment or post submissions. Thus the average user is content just reading the submissions.

In the future, models with better predictive ability could be developed. Prior work in the software engineering domain sentiment analysis \cite{obaidi2022sentiment} suggests that using BERT\cite{devlin2018bert} could lead to more accurate models. However, the limitation of these pre-trained deep-learning models is their inherent black-box nature. While they could improve the accuracy of humor prediction, they would not provide humanly understandable predictions, unlike the LDA and unigram models that we have used.

%% file: conclusion.tex
\section{Conclusion}\label{sec:conclusion}
This study investigated a novel topic in software engineering literature: humor used by users in a software development-related community on the social media site Reddit. We built prediction models and found that predicting humor is difficult, with the best model explaining only 10\% of the variance. However, we did identify meaningful predictors for humor scores. Our analysis indicates that the highest submission scores are typically achieved by image-based submissions crafted during the winter months of the northern hemisphere, between 2-3pm UTC on weekends. These submissions often embody themes rooted in superiority theory, where we feel superior to the object we are laughing at, and incongruity theory, where humor arises from a discrepancy between what is expected and what actually occurs, leading to a moment of surprise or cognitive dissonance that we find amusing. Regression with topic modeling revealed that the topics of Learning, Writing Code, and Fixing Bugs were the best at predicting submission humor scores. 

\section*{Data Availability}
\label{sec:data_availability}
We share the dataset with topic information and the dataset with randomly sampled human-labeled data in our replication package on Zenodo~\cite{zenodo_anon}. The replication package also includes a graph showing the perplexity score across different numbers of topics, the scripts used for data analysis, and the Python script used for clustering.

\section*{Ethics Statement}
\label{sec:ethics}
To mitigate concerns about tying specific submissions to individuals who wish to remain anonymous, we have removed the API-provided identifiers from the shared dataset. Furthermore, the users referred to in Section \ref{sec:RQ1} have audiences ranging from hundreds of thousands to millions.

\begin{acks}\label{sec:ack}
The first author has been funded by the Killam Postdoctoral Fellowship. The first, third, fourth, and fifth authors have been funded by the Strategic Research Council of Research Council of Finland (Grant IDs 358471 and 243046891).
\end{acks}

%% file: main.bbl

\begin{thebibliography}{49}


\ifx \showCODEN    \undefined \def \showCODEN     #1{\unskip}     \fi
\ifx \showDOI      \undefined \def \showDOI       #1{#1}\fi
\ifx \showISBNx    \undefined \def \showISBNx     #1{\unskip}     \fi
\ifx \showISBNxiii \undefined \def \showISBNxiii  #1{\unskip}     \fi
\ifx \showISSN     \undefined \def \showISSN      #1{\unskip}     \fi
\ifx \showLCCN     \undefined \def \showLCCN      #1{\unskip}     \fi
\ifx \shownote     \undefined \def \shownote      #1{#1}          \fi
\ifx \showarticletitle \undefined \def \showarticletitle #1{#1}   \fi
\ifx \showURL      \undefined \def \showURL       {\relax}        \fi
\providecommand\bibfield[2]{#2}
\providecommand\bibinfo[2]{#2}
\providecommand\natexlab[1]{#1}
\providecommand\showeprint[2][]{arXiv:#2}

\bibitem[Bardon(2005)]%
        {bardon2005philosophy}
\bibfield{author}{\bibinfo{person}{Adrian Bardon}.} \bibinfo{year}{2005}\natexlab{}.
\newblock \showarticletitle{The philosophy of humor}.
\newblock \bibinfo{journal}{\emph{Comedy: A geographic and historical guide}}  \bibinfo{volume}{2} (\bibinfo{year}{2005}), \bibinfo{pages}{462--476}.
\newblock


\bibitem[Baumgartner et~al\mbox{.}(2020)]%
        {baumgartner2020pushshift}
\bibfield{author}{\bibinfo{person}{Jason Baumgartner}, \bibinfo{person}{Savvas Zannettou}, \bibinfo{person}{Brian Keegan}, \bibinfo{person}{Megan Squire}, {and} \bibinfo{person}{Jeremy Blackburn}.} \bibinfo{year}{2020}\natexlab{}.
\newblock \showarticletitle{The pushshift reddit dataset}. In \bibinfo{booktitle}{\emph{Proceedings of the international AAAI conference on web and social media}}, Vol.~\bibinfo{volume}{14}. \bibinfo{pages}{830--839}.
\newblock


\bibitem[Bird et~al\mbox{.}(2009)]%
        {bird2009natural}
\bibfield{author}{\bibinfo{person}{Steven Bird}, \bibinfo{person}{Ewan Klein}, {and} \bibinfo{person}{Edward Loper}.} \bibinfo{year}{2009}\natexlab{}.
\newblock \bibinfo{booktitle}{\emph{Natural language processing with Python: analyzing text with the natural language toolkit}}.
\newblock \bibinfo{publisher}{" O'Reilly Media, Inc."}.
\newblock


\bibitem[Blei et~al\mbox{.}(2003)]%
        {blei2003latent}
\bibfield{author}{\bibinfo{person}{David~M Blei}, \bibinfo{person}{Andrew~Y Ng}, {and} \bibinfo{person}{Michael~I Jordan}.} \bibinfo{year}{2003}\natexlab{}.
\newblock \showarticletitle{Latent dirichlet allocation}.
\newblock \bibinfo{journal}{\emph{Journal of machine Learning research}} \bibinfo{volume}{3}, \bibinfo{number}{Jan} (\bibinfo{year}{2003}), \bibinfo{pages}{993--1022}.
\newblock


\bibitem[Borsotti and Bj{\o}rn(2022)]%
        {borsotti2022humor}
\bibfield{author}{\bibinfo{person}{Valeria Borsotti} {and} \bibinfo{person}{Pernille Bj{\o}rn}.} \bibinfo{year}{2022}\natexlab{}.
\newblock \showarticletitle{Humor and stereotypes in computing: An equity-focused approach to institutional accountability}.
\newblock \bibinfo{journal}{\emph{Computer Supported Cooperative Work (CSCW)}} \bibinfo{volume}{31}, \bibinfo{number}{4} (\bibinfo{year}{2022}), \bibinfo{pages}{771--803}.
\newblock


\bibitem[Churchill and Singh(2022)]%
        {churchill2022evolution}
\bibfield{author}{\bibinfo{person}{Rob Churchill} {and} \bibinfo{person}{Lisa Singh}.} \bibinfo{year}{2022}\natexlab{}.
\newblock \showarticletitle{The evolution of topic modeling}.
\newblock \bibinfo{journal}{\emph{Comput. Surveys}} \bibinfo{volume}{54}, \bibinfo{number}{10s} (\bibinfo{year}{2022}), \bibinfo{pages}{1--35}.
\newblock


\bibitem[Claes et~al\mbox{.}(2018)]%
        {claes2018programmers}
\bibfield{author}{\bibinfo{person}{Ma{\"e}lick Claes}, \bibinfo{person}{Mika~V M{\"a}ntyl{\"a}}, \bibinfo{person}{Miikka Kuutila}, {and} \bibinfo{person}{Bram Adams}.} \bibinfo{year}{2018}\natexlab{}.
\newblock \showarticletitle{Do programmers work at night or during the weekend?}. In \bibinfo{booktitle}{\emph{Proceedings of the 40th International Conference on Software Engineering}}. \bibinfo{pages}{705--715}.
\newblock


\bibitem[Devlin et~al\mbox{.}(2019)]%
        {devlin2018bert}
\bibfield{author}{\bibinfo{person}{Jacob Devlin}, \bibinfo{person}{Chang Ming-Wei}, \bibinfo{person}{Lee Kenton}, {and} \bibinfo{person}{Kristina Toutanova}.} \bibinfo{year}{2019}\natexlab{}.
\newblock \showarticletitle{BERT: Pre-training of Deep Bidirectional Transformers for Language Understanding}. In \bibinfo{booktitle}{\emph{Proceedings of NAACL-HLT}}. \bibinfo{pages}{4171--4186}.
\newblock


\bibitem[Duncan et~al\mbox{.}(1990)]%
        {duncan1990humor}
\bibfield{author}{\bibinfo{person}{W~Jack Duncan}, \bibinfo{person}{Larry~R Smeltzer}, {and} \bibinfo{person}{Terry~L Leap}.} \bibinfo{year}{1990}\natexlab{}.
\newblock \showarticletitle{Humor and work: Applications of joking behavior to management}.
\newblock \bibinfo{journal}{\emph{Journal of Management}} \bibinfo{volume}{16}, \bibinfo{number}{2} (\bibinfo{year}{1990}), \bibinfo{pages}{255--278}.
\newblock


\bibitem[Fan et~al\mbox{.}(2020)]%
        {FAN2020105}
\bibfield{author}{\bibinfo{person}{Xiaochao Fan}, \bibinfo{person}{Hongfei Lin}, \bibinfo{person}{Liang Yang}, \bibinfo{person}{Yufeng Diao}, \bibinfo{person}{Chen Shen}, \bibinfo{person}{Yonghe Chu}, {and} \bibinfo{person}{Yanbo Zou}.} \bibinfo{year}{2020}\natexlab{}.
\newblock \showarticletitle{Humor detection via an internal and external neural network}.
\newblock \bibinfo{journal}{\emph{Neurocomputing}}  \bibinfo{volume}{394} (\bibinfo{year}{2020}), \bibinfo{pages}{105--111}.
\newblock
\showISSN{0925-2312}
\urldef\tempurl%
\url{https://doi.org/10.1016/j.neucom.2020.02.030}
\showDOI{\tempurl}


\bibitem[Feinerer et~al\mbox{.}(2008)]%
        {feinerer2008text}
\bibfield{author}{\bibinfo{person}{Ingo Feinerer}, \bibinfo{person}{Kurt Hornik}, {and} \bibinfo{person}{David Meyer}.} \bibinfo{year}{2008}\natexlab{}.
\newblock \showarticletitle{Text mining infrastructure in R}.
\newblock \bibinfo{journal}{\emph{Journal of statistical software}}  \bibinfo{volume}{25} (\bibinfo{year}{2008}), \bibinfo{pages}{1--54}.
\newblock


\bibitem[Ford et~al\mbox{.}(2015)]%
        {ford2015disparagement}
\bibfield{author}{\bibinfo{person}{Thomas~E Ford}, \bibinfo{person}{Kyle Richardson}, {and} \bibinfo{person}{Whitney~E Petit}.} \bibinfo{year}{2015}\natexlab{}.
\newblock \showarticletitle{Disparagement humor and prejudice: Contemporary theory and research}.
\newblock \bibinfo{journal}{\emph{Humor}} \bibinfo{volume}{28}, \bibinfo{number}{2} (\bibinfo{year}{2015}), \bibinfo{pages}{171--186}.
\newblock


\bibitem[Friedman et~al\mbox{.}(2021)]%
        {friedman2021package}
\bibfield{author}{\bibinfo{person}{Jerome Friedman}, \bibinfo{person}{Trevor Hastie}, \bibinfo{person}{Rob Tibshirani}, \bibinfo{person}{Balasubramanian Narasimhan}, \bibinfo{person}{Kenneth Tay}, \bibinfo{person}{Noah Simon}, {and} \bibinfo{person}{Junyang Qian}.} \bibinfo{year}{2021}\natexlab{}.
\newblock \showarticletitle{Package ‘glmnet’}.
\newblock \bibinfo{journal}{\emph{CRAN R Repositary}} (\bibinfo{year}{2021}).
\newblock


\bibitem[Friedman and Friedman(2002)]%
        {friedman2002computer}
\bibfield{author}{\bibinfo{person}{Linda~Weiser Friedman} {and} \bibinfo{person}{Hershey~H Friedman}.} \bibinfo{year}{2002}\natexlab{}.
\newblock \bibinfo{booktitle}{\emph{Computer-Oriented HUMor (COHUM):‘I get it.’}}.
\newblock \bibinfo{type}{{T}echnical {R}eport}. \bibinfo{institution}{Working Paper\# CIS-2002-10, CIS Working Paper Series, Zicklin School of Baruch College, City University of New York, New York, NY, http://cisnet.baruch.cuny.edu/papers/cis200210.html}.
\newblock


\bibitem[Friedman and Friedman(2003)]%
        {friedman2003framework}
\bibfield{author}{\bibinfo{person}{Linda~Weiser Friedman} {and} \bibinfo{person}{Hershey~H Friedman}.} \bibinfo{year}{2003}\natexlab{}.
\newblock \bibinfo{booktitle}{\emph{A Framework for the study of computer-oriented humor (Cohum)}}.
\newblock \bibinfo{type}{{T}echnical {R}eport}. \bibinfo{institution}{CIS Working Paper}.
\newblock


\bibitem[Gamer et~al\mbox{.}(2012)]%
        {gamer2012package}
\bibfield{author}{\bibinfo{person}{Matthias Gamer}, \bibinfo{person}{Jim Lemon}, \bibinfo{person}{Maintainer~Matthias Gamer}, \bibinfo{person}{A Robinson}, {and} \bibinfo{person}{W Kendall’s}.} \bibinfo{year}{2012}\natexlab{}.
\newblock \showarticletitle{Package ‘irr’}.
\newblock \bibinfo{journal}{\emph{Various coefficients of interrater reliability and agreement}}  \bibinfo{volume}{22} (\bibinfo{year}{2012}), \bibinfo{pages}{1--32}.
\newblock


\bibitem[Graziotin et~al\mbox{.}(2014)]%
        {graziotin2014happy}
\bibfield{author}{\bibinfo{person}{Daniel Graziotin}, \bibinfo{person}{Xiaofeng Wang}, {and} \bibinfo{person}{Pekka Abrahamsson}.} \bibinfo{year}{2014}\natexlab{}.
\newblock \showarticletitle{Happy software developers solve problems better: psychological measurements in empirical software engineering}.
\newblock \bibinfo{journal}{\emph{PeerJ}}  \bibinfo{volume}{2} (\bibinfo{year}{2014}), \bibinfo{pages}{e289}.
\newblock


\bibitem[Hellman et~al\mbox{.}(2022)]%
        {hellman2022characterizing}
\bibfield{author}{\bibinfo{person}{Jazlyn Hellman}, \bibinfo{person}{Jiahao Chen}, \bibinfo{person}{Md~Sami Uddin}, \bibinfo{person}{Jinghui Cheng}, {and} \bibinfo{person}{Jin~LC Guo}.} \bibinfo{year}{2022}\natexlab{}.
\newblock \showarticletitle{Characterizing user behaviors in open-source software user forums: an empirical study}. In \bibinfo{booktitle}{\emph{Proceedings of the 15th International Conference on Cooperative and Human Aspects of Software Engineering}}. \bibinfo{pages}{46--55}.
\newblock


\bibitem[Hong and Davison(2010)]%
        {hong2010empirical}
\bibfield{author}{\bibinfo{person}{Liangjie Hong} {and} \bibinfo{person}{Brian~D Davison}.} \bibinfo{year}{2010}\natexlab{}.
\newblock \showarticletitle{Empirical study of topic modeling in twitter}. In \bibinfo{booktitle}{\emph{Proceedings of the first workshop on social media analytics}}. \bibinfo{pages}{80--88}.
\newblock


\bibitem[Huang(2022)]%
        {huang2022leader}
\bibfield{author}{\bibinfo{person}{Mei-Jun Huang}.} \bibinfo{year}{2022}\natexlab{}.
\newblock \showarticletitle{Leader self-deprecating humor and employee creativity at workplace: a longitudinal study}.
\newblock \bibinfo{journal}{\emph{Review of Managerial Science}} (\bibinfo{year}{2022}), \bibinfo{pages}{1--26}.
\newblock


\bibitem[Huang et~al\mbox{.}(2018)]%
        {huang2018identifying}
\bibfield{author}{\bibinfo{person}{Qiao Huang}, \bibinfo{person}{Emad Shihab}, \bibinfo{person}{Xin Xia}, \bibinfo{person}{David Lo}, {and} \bibinfo{person}{Shanping Li}.} \bibinfo{year}{2018}\natexlab{}.
\newblock \showarticletitle{Identifying self-admitted technical debt in open source projects using text mining}.
\newblock \bibinfo{journal}{\emph{Empirical Software Engineering}} \bibinfo{volume}{23}, \bibinfo{number}{1} (\bibinfo{year}{2018}), \bibinfo{pages}{418--451}.
\newblock


\bibitem[Jeong et~al\mbox{.}(2019)]%
        {JEONG2019280}
\bibfield{author}{\bibinfo{person}{Byeongki Jeong}, \bibinfo{person}{Janghyeok Yoon}, {and} \bibinfo{person}{Jae-Min Lee}.} \bibinfo{year}{2019}\natexlab{}.
\newblock \showarticletitle{Social media mining for product planning: A product opportunity mining approach based on topic modeling and sentiment analysis}.
\newblock \bibinfo{journal}{\emph{International Journal of Information Management}}  \bibinfo{volume}{48} (\bibinfo{year}{2019}), \bibinfo{pages}{280--290}.
\newblock
\showISSN{0268-4012}
\urldef\tempurl%
\url{https://doi.org/10.1016/j.ijinfomgt.2017.09.009}
\showDOI{\tempurl}


\bibitem[Kimata(2004)]%
        {kimata2004laughter}
\bibfield{author}{\bibinfo{person}{Hajime Kimata}.} \bibinfo{year}{2004}\natexlab{}.
\newblock \showarticletitle{Laughter Counteracts Enhancement of Plasma Neurotrophin Levels and Allergic Skin Wheal Responses by Mobile Phone—Mediated Stress}.
\newblock \bibinfo{journal}{\emph{Behavioral Medicine}} \bibinfo{volume}{29}, \bibinfo{number}{4} (\bibinfo{year}{2004}), \bibinfo{pages}{149--154}.
\newblock


\bibitem[Knegtmans et~al\mbox{.}(2018)]%
        {knegtmans2018impact}
\bibfield{author}{\bibinfo{person}{Hans Knegtmans}, \bibinfo{person}{Wilco~W Van~Dijk}, \bibinfo{person}{Marlon Mooijman}, \bibinfo{person}{Nina Van~Lier}, \bibinfo{person}{Sacha Rintjema}, {and} \bibinfo{person}{Annemieke Wassink}.} \bibinfo{year}{2018}\natexlab{}.
\newblock \showarticletitle{The impact of social power on the evaluation of offensive jokes}.
\newblock \bibinfo{journal}{\emph{Humor}} \bibinfo{volume}{31}, \bibinfo{number}{1} (\bibinfo{year}{2018}), \bibinfo{pages}{85--104}.
\newblock


\bibitem[Kuutila et~al\mbox{.}(2021)]%
        {kuutila2021individual}
\bibfield{author}{\bibinfo{person}{Miikka Kuutila}, \bibinfo{person}{Mika M{\"a}ntyl{\"a}}, \bibinfo{person}{Ma{\"e}lick Claes}, \bibinfo{person}{Marko Elovainio}, {and} \bibinfo{person}{Bram Adams}.} \bibinfo{year}{2021}\natexlab{}.
\newblock \showarticletitle{Individual differences limit predicting well-being and productivity using software repositories: a longitudinal industrial study}.
\newblock \bibinfo{journal}{\emph{Empirical Software Engineering}} \bibinfo{volume}{26}, \bibinfo{number}{5} (\bibinfo{year}{2021}), \bibinfo{pages}{88}.
\newblock


\bibitem[Kuutila et~al\mbox{.}(2020)]%
        {kuutila2020chat}
\bibfield{author}{\bibinfo{person}{Miikka Kuutila}, \bibinfo{person}{Mika~V M{\~a}ntyl{\~a}}, {and} \bibinfo{person}{Ma{\"e}lick Claes}.} \bibinfo{year}{2020}\natexlab{}.
\newblock \showarticletitle{Chat activity is a better predictor than chat sentiment on software developers productivity}. In \bibinfo{booktitle}{\emph{Proceedings of the IEEE/ACM 42nd international conference on software engineering workshops}}. \bibinfo{pages}{553--556}.
\newblock


\bibitem[Kuutila et~al\mbox{.}(2024)]%
        {zenodo_anon}
\bibfield{author}{\bibinfo{person}{Miikka Kuutila}, \bibinfo{person}{Leevi Rantala}, \bibinfo{person}{Junhao Li}, \bibinfo{person}{Simo Hosio}, {and} \bibinfo{person}{Mika M{\"a}ntyl{\"a}}.} \bibinfo{year}{2024}\natexlab{}.
\newblock \bibinfo{title}{Replication package for the paper "What Makes Programmers Laugh? Exploring the Submissions of the Subreddit r/ProgrammerHumor.}
\newblock
\newblock
\urldef\tempurl%
\url{https://doi.org/10.5281/zenodo.11124117}
\showDOI{\tempurl}
\newblock
\shownote{DOI =10.5281/zenodo.11124117, URL = https://doi.org/10.5281/zenodo.11124117}.


\bibitem[Landis and Koch(1977)]%
        {landis1977measurement}
\bibfield{author}{\bibinfo{person}{J~Richard Landis} {and} \bibinfo{person}{Gary~G Koch}.} \bibinfo{year}{1977}\natexlab{}.
\newblock \showarticletitle{The measurement of observer agreement for categorical data}.
\newblock \bibinfo{journal}{\emph{biometrics}} (\bibinfo{year}{1977}), \bibinfo{pages}{159--174}.
\newblock


\bibitem[Li et~al\mbox{.}(2020)]%
        {LI2020102290}
\bibfield{author}{\bibinfo{person}{Da Li}, \bibinfo{person}{Rafal Rzepka}, \bibinfo{person}{Michal Ptaszynski}, {and} \bibinfo{person}{Kenji Araki}.} \bibinfo{year}{2020}\natexlab{}.
\newblock \showarticletitle{HEMOS: A novel deep learning-based fine-grained humor detecting method for sentiment analysis of social media}.
\newblock \bibinfo{journal}{\emph{Information Processing \& Management}} \bibinfo{volume}{57}, \bibinfo{number}{6} (\bibinfo{year}{2020}), \bibinfo{pages}{102290}.
\newblock
\showISSN{0306-4573}
\urldef\tempurl%
\url{https://doi.org/10.1016/j.ipm.2020.102290}
\showDOI{\tempurl}


\bibitem[Lintott(2016)]%
        {lintott2016superiority}
\bibfield{author}{\bibinfo{person}{Sheila Lintott}.} \bibinfo{year}{2016}\natexlab{}.
\newblock \showarticletitle{Superiority in humor theory}.
\newblock \bibinfo{journal}{\emph{The Journal of Aesthetics and Art Criticism}} \bibinfo{volume}{74}, \bibinfo{number}{4} (\bibinfo{year}{2016}), \bibinfo{pages}{347--358}.
\newblock


\bibitem[Liu et~al\mbox{.}(2021)]%
        {liu2021examining}
\bibfield{author}{\bibinfo{person}{Xia Liu}, \bibinfo{person}{Hyunju Shin}, {and} \bibinfo{person}{Alvin~C Burns}.} \bibinfo{year}{2021}\natexlab{}.
\newblock \showarticletitle{Examining the impact of luxury brand's social media marketing on customer engagement: Using big data analytics and natural language processing}.
\newblock \bibinfo{journal}{\emph{Journal of Business Research}}  \bibinfo{volume}{125} (\bibinfo{year}{2021}), \bibinfo{pages}{815--826}.
\newblock


\bibitem[Mahadi et~al\mbox{.}(2022)]%
        {mahadi2022conclusion}
\bibfield{author}{\bibinfo{person}{Alvi Mahadi}, \bibinfo{person}{Neil~A Ernst}, {and} \bibinfo{person}{Karan Tongay}.} \bibinfo{year}{2022}\natexlab{}.
\newblock \showarticletitle{Conclusion stability for natural language based mining of design discussions}.
\newblock \bibinfo{journal}{\emph{Empirical Software Engineering}} \bibinfo{volume}{27}, \bibinfo{number}{1} (\bibinfo{year}{2022}), \bibinfo{pages}{1--42}.
\newblock


\bibitem[Mesmer-Magnus et~al\mbox{.}(2012)]%
        {mesmer2012meta}
\bibfield{author}{\bibinfo{person}{Jessica Mesmer-Magnus}, \bibinfo{person}{David~J Glew}, {and} \bibinfo{person}{Chockalingam Viswesvaran}.} \bibinfo{year}{2012}\natexlab{}.
\newblock \showarticletitle{A meta-analysis of positive humor in the workplace}.
\newblock \bibinfo{journal}{\emph{Journal of Managerial Psychology}}  \bibinfo{volume}{27} (\bibinfo{year}{2012}), \bibinfo{pages}{155--190}.
\newblock


\bibitem[Morales and Zhai(2017)]%
        {morales2017identifying}
\bibfield{author}{\bibinfo{person}{Alex Morales} {and} \bibinfo{person}{ChengXiang Zhai}.} \bibinfo{year}{2017}\natexlab{}.
\newblock \showarticletitle{Identifying humor in reviews using background text sources}. In \bibinfo{booktitle}{\emph{Proceedings of the 2017 Conference on Empirical Methods in Natural Language Processing}}. \bibinfo{pages}{492--501}.
\newblock


\bibitem[Moster et~al\mbox{.}(2022)]%
        {moster2022zone}
\bibfield{author}{\bibinfo{person}{Makayla Moster}, \bibinfo{person}{Aarav Chandra}, \bibinfo{person}{Christal Chu}, \bibinfo{person}{Weiyi Liu}, {and} \bibinfo{person}{Paige Rodeghero}.} \bibinfo{year}{2022}\natexlab{}.
\newblock \showarticletitle{In the zone: An analysis of the music practices of remote software developers}. In \bibinfo{booktitle}{\emph{Proceedings of the 16th ACM/IEEE International Symposium on Empirical Software Engineering and Measurement}}. \bibinfo{pages}{313--318}.
\newblock


\bibitem[Nguyen and Shirai(2015)]%
        {nguyen2015topic}
\bibfield{author}{\bibinfo{person}{Thien~Hai Nguyen} {and} \bibinfo{person}{Kiyoaki Shirai}.} \bibinfo{year}{2015}\natexlab{}.
\newblock \showarticletitle{Topic modeling based sentiment analysis on social media for stock market prediction}. In \bibinfo{booktitle}{\emph{Proceedings of the 53rd Annual Meeting of the Association for Computational Linguistics and the 7th International Joint Conference on Natural Language Processing (Volume 1: Long Papers)}}. \bibinfo{pages}{1354--1364}.
\newblock


\bibitem[Obaidi and Kl{\"u}nder(2021)]%
        {obaidi2021development}
\bibfield{author}{\bibinfo{person}{Martin Obaidi} {and} \bibinfo{person}{Jil Kl{\"u}nder}.} \bibinfo{year}{2021}\natexlab{}.
\newblock \showarticletitle{Development and application of sentiment analysis tools in software engineering: A systematic literature review}. In \bibinfo{booktitle}{\emph{Proceedings of the 25th International Conference on Evaluation and Assessment in Software Engineering}}. \bibinfo{pages}{80--89}.
\newblock


\bibitem[Obaidi et~al\mbox{.}(2022)]%
        {obaidi2022sentiment}
\bibfield{author}{\bibinfo{person}{Martin Obaidi}, \bibinfo{person}{Lukas Nagel}, \bibinfo{person}{Alexander Specht}, {and} \bibinfo{person}{Jil Kl{\"u}nder}.} \bibinfo{year}{2022}\natexlab{}.
\newblock \showarticletitle{Sentiment analysis tools in software engineering: A systematic mapping study}.
\newblock \bibinfo{journal}{\emph{Information and software Technology}}  \bibinfo{volume}{151} (\bibinfo{year}{2022}), \bibinfo{pages}{107018}.
\newblock


\bibitem[Pennacchiotti and Gurumurthy(2011)]%
        {pennacchiotti2011investigating}
\bibfield{author}{\bibinfo{person}{Marco Pennacchiotti} {and} \bibinfo{person}{Siva Gurumurthy}.} \bibinfo{year}{2011}\natexlab{}.
\newblock \showarticletitle{Investigating topic models for social media user recommendation}. In \bibinfo{booktitle}{\emph{Proceedings of the 20th international conference companion on World wide web}}. \bibinfo{pages}{101--102}.
\newblock


\bibitem[Radford et~al\mbox{.}(2019)]%
        {radford2019language}
\bibfield{author}{\bibinfo{person}{Alec Radford}, \bibinfo{person}{Jeffrey Wu}, \bibinfo{person}{Rewon Child}, \bibinfo{person}{David Luan}, \bibinfo{person}{Dario Amodei}, \bibinfo{person}{Ilya Sutskever}, {et~al\mbox{.}}} \bibinfo{year}{2019}\natexlab{}.
\newblock \showarticletitle{Language models are unsupervised multitask learners}.
\newblock \bibinfo{journal}{\emph{OpenAI blog}} \bibinfo{volume}{1}, \bibinfo{number}{8} (\bibinfo{year}{2019}), \bibinfo{pages}{9}.
\newblock


\bibitem[Ritchie(2004)]%
        {ritchie2004linguistic}
\bibfield{author}{\bibinfo{person}{Graeme Ritchie}.} \bibinfo{year}{2004}\natexlab{}.
\newblock \bibinfo{booktitle}{\emph{The linguistic analysis of jokes}}.
\newblock \bibinfo{publisher}{Routledge}.
\newblock


\bibitem[Sacco et~al\mbox{.}(2021)]%
        {sacco2021not}
\bibfield{author}{\bibinfo{person}{Donald~F Sacco}, \bibinfo{person}{Mitch Brown}, {and} \bibinfo{person}{Haley~D May}.} \bibinfo{year}{2021}\natexlab{}.
\newblock \showarticletitle{Not taking a joke: the influence of target status, sex, and age on reactions to workplace humor}.
\newblock \bibinfo{journal}{\emph{Psychological Reports}} \bibinfo{volume}{124}, \bibinfo{number}{3} (\bibinfo{year}{2021}), \bibinfo{pages}{1316--1334}.
\newblock


\bibitem[Sarker et~al\mbox{.}(2023)]%
        {sarker2023toxispanse}
\bibfield{author}{\bibinfo{person}{Jaydeb Sarker}, \bibinfo{person}{Sayma Sultana}, \bibinfo{person}{Steven~R Wilson}, {and} \bibinfo{person}{Amiangshu Bosu}.} \bibinfo{year}{2023}\natexlab{}.
\newblock \showarticletitle{ToxiSpanSE: An Explainable Toxicity Detection in Code Review Comments}. In \bibinfo{booktitle}{\emph{2023 ACM/IEEE International Symposium on Empirical Software Engineering and Measurement (ESEM)}}. IEEE, \bibinfo{pages}{1--12}.
\newblock


\bibitem[Shaw(2010)]%
        {shaw2010philosophy}
\bibfield{author}{\bibinfo{person}{Joshua Shaw}.} \bibinfo{year}{2010}\natexlab{}.
\newblock \showarticletitle{Philosophy of humor}.
\newblock \bibinfo{journal}{\emph{Philosophy Compass}} \bibinfo{volume}{5}, \bibinfo{number}{2} (\bibinfo{year}{2010}), \bibinfo{pages}{112--126}.
\newblock


\bibitem[Sultana et~al\mbox{.}(2021)]%
        {sultana2021rubric}
\bibfield{author}{\bibinfo{person}{Sayma Sultana}, \bibinfo{person}{Jaydeb Sarker}, {and} \bibinfo{person}{Amiangshu Bosu}.} \bibinfo{year}{2021}\natexlab{}.
\newblock \showarticletitle{A Rubric to Identify Misogynistic and Sexist Texts from Software Developer Communications}. In \bibinfo{booktitle}{\emph{Proceedings of the 15th ACM/IEEE International Symposium on Empirical Software Engineering and Measurement (ESEM)}}. \bibinfo{pages}{1--6}.
\newblock


\bibitem[Thompson and Mimno(2020)]%
        {thompson2020topic}
\bibfield{author}{\bibinfo{person}{Laure Thompson} {and} \bibinfo{person}{David Mimno}.} \bibinfo{year}{2020}\natexlab{}.
\newblock \showarticletitle{Topic modeling with contextualized word representation clusters}.
\newblock \bibinfo{journal}{\emph{arXiv preprint arXiv:2010.12626}} (\bibinfo{year}{2020}).
\newblock


\bibitem[Weissburg et~al\mbox{.}(2022)]%
        {weissburg2022judging}
\bibfield{author}{\bibinfo{person}{Evan Weissburg}, \bibinfo{person}{Arya Kumar}, {and} \bibinfo{person}{Paramveer~S Dhillon}.} \bibinfo{year}{2022}\natexlab{}.
\newblock \showarticletitle{Judging a book by its cover: Predicting the marginal impact of title on Reddit post popularity}. In \bibinfo{booktitle}{\emph{Proceedings of the International AAAI Conference on Web and Social Media}}, Vol.~\bibinfo{volume}{16}. \bibinfo{pages}{1098--1108}.
\newblock


\bibitem[Weninger et~al\mbox{.}(2015)]%
        {weninger2015random}
\bibfield{author}{\bibinfo{person}{Tim Weninger}, \bibinfo{person}{Thomas~James Johnston}, {and} \bibinfo{person}{Maria Glenski}.} \bibinfo{year}{2015}\natexlab{}.
\newblock \showarticletitle{Random voting effects in social-digital spaces: A case study of reddit post submissions}. In \bibinfo{booktitle}{\emph{Proceedings of the 26th ACM conference on hypertext \& social media}}. \bibinfo{pages}{293--297}.
\newblock


\bibitem[Wilkins and Eisenbraun(2009)]%
        {wilkins2009humor}
\bibfield{author}{\bibinfo{person}{Julia Wilkins} {and} \bibinfo{person}{Amy~Janel Eisenbraun}.} \bibinfo{year}{2009}\natexlab{}.
\newblock \showarticletitle{Humor theories and the physiological benefits of laughter}.
\newblock \bibinfo{journal}{\emph{Holistic nursing practice}} \bibinfo{volume}{23}, \bibinfo{number}{6} (\bibinfo{year}{2009}), \bibinfo{pages}{349--354}.
\newblock


\end{thebibliography}
